\documentclass[a4paper]{article}

\usepackage[utf8]{inputenc}
\usepackage[T1]{fontenc}
\usepackage{multirow}
\usepackage{tabularx}
\usepackage{float}
\usepackage{xcolor}

\usepackage{INTERSPEECH2020}

\usepackage{enumitem} 
\usepackage[square,numbers]{natbib} 
\usepackage[italic]{mathastext} 
\usepackage[hidelinks]{hyperref} 
\usepackage{microtype}  

\newcommand\Tstrut{\rule{0pt}{2.2ex}}

\title{Single Model, Many Languages: Meta-learning for Truly Multilingual TTS}
\title{One Model, Many Languages: Meta-learning for Multilingual Text-to-Speech}
\name{Tomáš Nekvinda, Ondřej Dušek}
\address{
  Charles University, Faculty of Mathematics and Physics, Prague, Czechia}
\email{tom@neqindi.cz, odusek@ufal.mff.cuni.cz}

\usepackage[normalem]{ulem}

\def\ODdel{\bgroup\markoverwith{\textcolor{cyan!89!yellow!80!black!100}{\rule[0.4ex]{2pt}{3pt}}}\ULon}
\MakeRobust\ODdel

\usepackage[normalem]{ulem}

\def\TNdel{\bgroup\markoverwith{\textcolor{magenta!100!yellow!45!black!100}{\rule[0.4ex]{2pt}{3pt}}}\ULon}
\MakeRobust\TNdel

\makeatletter
\renewcommand{\paragraph}{%
  \@startsection{paragraph}{4}%
  {\z@}{0.6ex \@plus 1ex \@minus 0.2ex}{-1em}%
  {\normalfont\normalsize\bfseries}%
}
\makeatother

\usepackage{xspace}
\def\cosi#1{\textsc{#1}} 
\def\gen{\cosi{Gen}\xspace}
\def\sep{\cosi{Sep}\xspace}
\def\sha{\cosi{Sha}\xspace}
\def\sgl{\cosi{Sgl}\xspace}

\begin{document}

\maketitle
%

\begin{abstract}
  We introduce an approach to multilingual speech synthesis which uses the meta-learning concept of contextual parameter generation and produces natural-sounding multilingual speech using more languages and less training data than previous approaches. Our model is based on Tacotron~2 with a fully convolutional input text encoder whose weights are predicted by a separate parameter generator network. To boost voice cloning, the model uses an adversarial speaker classifier with a gradient reversal layer that removes speaker-specific information from the encoder.
  
  We arranged two experiments to compare our model with baselines using various levels of cross-lingual parameter sharing, in order to evaluate: (1) stability and performance when training on low amounts of data, (2) pronunciation accuracy and voice quality of code-switching synthesis. For training, we used the CSS10 dataset and our new small dataset based on Common Voice recordings in five languages. Our model is shown to effectively share information across languages and according to a subjective evaluation test, it produces more natural and accurate code-switching speech than the baselines.
  

  

\end{abstract}

\section{Introduction}

Contemporary end-to-end speech synthesis systems achieve great results and produce natural-sounding human-like speech \cite{shen18, oord16} even in real time \cite{kalchbrenner18, kumar19}. They make possible an efficient training that does not put high demands on quality, amount, and preprocessing of training data. Based on these advances, researchers aim at, for example, expressiveness \cite{wang18}, controllability \cite{hsu19}, or few-shot voice cloning \cite{jia2018}. When extending these models to support multiple languages, one may encounter obstacles such as different input representations or pronunciations, and imbalanced amounts of training data per language.

In this work, we examine cross-lingual knowledge-sharing aspects of multilingual text-to-speech (TTS). We experiment with more languages simultaneously than most previous TTS work 
known to us.
We can  summarize our contributions as follows:
(1) We propose a scalable grapheme-based model that utilizes the idea of contextual parameter generator network \cite{platanios18} and we compare it with baseline models using different levels of parameter sharing. 
(2) We introduce a new small dataset based on Common Voice \cite{ardila19} that includes data in five languages from 84 speakers.
(3) We evaluate effectiveness of the compared models on ten languages with three different scripts and we show their code-switching abilities on five languages. For the purposes of the evaluation, we created a new test set of 400 bilingual code-switching sentences.

\begin{figure}[t]
	\centering
	\includegraphics[width=\linewidth]{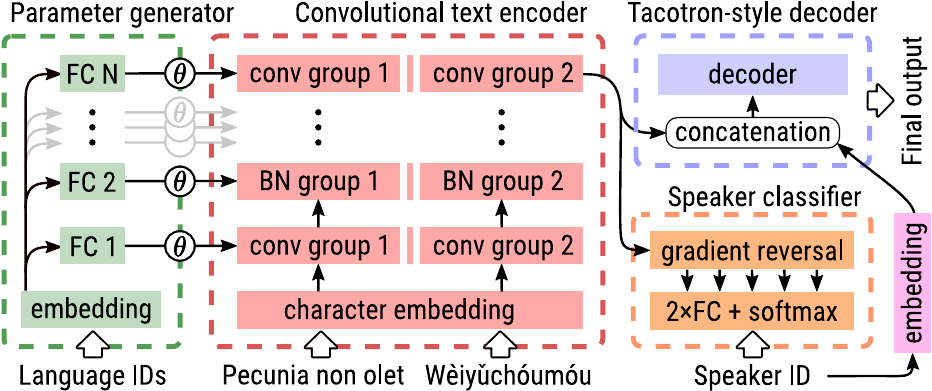}
	\caption{Diagram of our model. The meta-network generates parameters of language-specific convolutional text encoders. Encoded text inputs enhanced with speaker embeddings are read by the decoder. The adversarial classifier suppresses speaker-dependent information in encoder outputs.
	}
	\label{fig:generated}
\end{figure}

Our source code, hyper-parameters, training and evaluation data, samples, pre-trained models, and interactive demos are freely available on GitHub.\footnote{\url{https://github.com/Tomiinek/Multilingual\_Text\_to\_Speech} \label{fn:github}}

\section{Related Work}


So far, several works explored training joint multilingual models in text-to-speech, following similar experiments in the field of neural machine translation \cite{sachan18,platanios18}.
Multilingual models offer a few key benefits: 

\begin{itemize}[nosep,leftmargin=10pt]
\item \emph{Transfer learning:} We can try to make use of high-resource languages for training TTS systems
for low-resource languages, e.g., 
via transfer learning approaches \cite{jui19, lee18}. 

\item \emph{Knowledge sharing:} We may think of using multilingual data for joint training of a single shared text-to-speech model. Intuitively, this enables cross-lingual sharing of patterns learned from data. 
The only work in this area to our knowledge is \citeauthor{prakash2019}'s study \citep{prakash2019} on TTS for related Indian languages using hand-built unified phoneme representations.

\item \emph{Voice cloning:} 
Under certain circumstances, producing speech in multiple languages with the same voice, i.e., cross-lingual voice cloning, is desired. However, audio data where a single speaker speaks several languages is scarce. 
That is why multilingual voice-cloning systems should be trainable using mixtures of monolingual data. 
Here, \citet{zhang19} used Tacotron~2 \cite{shen18} conditioned on phonemes and showed voice-cloning abilities on English, Spanish, and Chinese. \citet{nachmani19} extended Voice Loop \cite{taigman18} and enabled voice conversion for English, Spanish, and German. \citet{Chen2019} used a phoneme-based Tacotron~2 with a ResCNN based speaker encoder \cite{li2017deep} that enables a massively multi-speaker speech synthesis, even with fictitious voices.

\item \emph{Code switching:} In this task closely related to cross-lingual voice cloning, we would like to alternate languages within sentences. This is useful for foreign names in navigation systems or news readers. In view of that, \citet{cao2019} modified Tacotron; their model uses language-specific encoders. Code-switching itself is done by combining of their outputs.
\end{itemize}
Overall, all recent multilingual text-to-speech systems were only tested in 2-3 languages simultaneously, or required vast amounts of data to be trained.

\section{Model Architecture}
\label{sec:models}


We base our experiments on Tacotron~2 \cite{shen18}. We focus on the spectrogram generation part here; for vocoding, we use WaveRNN \cite{kalchbrenner18, fatchord19} in all our configurations.
We first explain our new model that uses meta-learning for multilingual knowledge sharing in Sec.~\ref{sec:our-model}, then describe contrastive baseline models which are based on recent multilingual TTS architectures (Sec.~\ref{sec:baselines}).

\subsection{Our Model: Generated (\gen)}
\label{sec:our-model}

We introduce a scalable multilingual text-to-speech model that follows a meta-learning approach of contextual parameter generation proposed by \citet{platanios18} for NMT (see Fig.~\ref{fig:generated}). We call the model \emph{generated} (\gen) further in this text. 

The backbone of our model is built on our own implementation of Tacotron~2, composed of these main components: (1) an input text encoder that includes a stack of convolutional layers and a bidirectional LSTM, (2) a location-sensitive attention mechanism \cite{shen18} with the guided attention loss term \cite{tachibana17} that supports faster convergence, (3) a decoder with two stacked LSTM layers where the first queries the attention mechanism and the second generates outputs. We increase tolerance of the guided attention loss exponentially during training. 

We propose the following changes to this basic architecture:

\paragraph*{Convolutional Encoders:} We use multiple language-specific input text encoders. However, having a separate encoder with recurrent layers for each language is not practical
as it involves
passing the training batches (which 
should
be balanced with respect to languages) through multiple encoders sequentially. Therefore, we 
use a fully convolutional encoder from DCTTS \cite{tachibana17}. The encoders 
use
grouped layers and
are 
thus processed effectively. We enhance the encoders with batch normalization and dropout with a very low rate.
The normalization layers are situated before activations and dropouts after them.

\paragraph*{Encoder parameter generation:}
To enable cross-lingual knowledge-sharing, parameters of the encoders are generated using a separate network conditioned on language embeddings. 
The parameter generator is composed of multiple site-specific generators, each of which takes a language embedding on the input and produces parameters for one layer of the convolutional encoder for the given language. The generators enable a controllable cross-lingual parameter sharing because reduction of their size prevents generation of highly language-specific parameters. We implement them as fully connected layers.

\paragraph*{Training with multilingual batches:}
We construct unusual training batches to fully utilize the potential of this architecture. We would like to have a batch of $B$ examples that can be reshaped into a batch of size $B/L$ where $L$ is the number of encoder groups or languages. This new batch should have a new dimension that groups all examples with the same language. Thus we use a batch sampler that creates batches where for each $l < L$ and $i < B/L$, all $(l + iL)$-th examples are of the same language.

\paragraph*{Speaker embedding:}
We extend the model with a speaker embedding which is concatenated with each element of the encoded sequence that is attended by the decoder while generating spectrogram frames. This makes the model multi-speaker and allows cross-lingual voice cloning. 

\paragraph*{Adversarial speaker classifier:}
We combine the model with an adversarial speaker classifier \cite{zhang19} to boost voice cloning. The classifier follows principles of domain adversarial training \cite{ganin16} and is used to proactively remove speaker-specific information from the encoders. It includes a single hidden layer, a softmax layer, and a gradient reversal layer that scales the gradient flowing to the encoders by a factor $-\lambda$. The gradients are clipped to stabilize training. It is optimized to reduce the cross-entropy of speaker predictions. The predictions are done separately for each element of the encoders' outputs. 

\subsection{Baselines: Shared, Separate \& Single}
\label{sec:baselines}

We compare \gen with baseline models called \emph{shared} (\sha), \emph{separate} (\sep), and \emph{single} (\sgl). \sgl is a basic Tacotron~2 model, \sha and \sep follow the recent multilingual TTS works of \citet{zhang19} and \citet{cao2019}, respectively, but were slightly adapted to our tasks for a fairer comparison to \gen{} -- we use more languages and less data than the original works. In the following, we only describe their differences from \gen.

\paragraph*{Single (\sgl)} represents a set of monolingual models that follow vanilla Tacotron~2 \cite{shen18} with the original recurrent encoder and default settings. \sgl cannot be used for code-switching. 

\paragraph*{Shared (\sha):} Unlike \gen, \sha has a single encoder with the original Tacotron~2 architecture, so it fully shares all encoder parameters. This sharing implicitly leads to language-independent encoder outputs. The language-dependent processing happens in the decoder, so the speaker embeddings are explicitly factorized into speaker and language parts. 

\paragraph*{Separate (\sep)} uses multiple language-specific convolutional encoders too, but their parameters are not generated. It also does not include the adversarial speaker classifier.

\section{Dataset}
\label{sec:data}

We created a new dataset for our experiments, based on carefully cleaning and preprocessing freely available audio sources: CSS10 \cite{park19} and a small fraction of Common Voice \cite{ardila19}. Table~\ref{tab:data} shows total durations of the used audio data per language.

\begin{table}[t]
	\caption{Total data sizes per language (hours of audio data) in our cleaned CSS10 (CSS) and Common Voice (CV) subsets.}
	\label{tab:data}
	\centering

	\setlength{\tabcolsep}{3pt}
	\setlength{\extrarowheight}{0pt}
	\renewcommand{\arraystretch}{0.9}
	
	\begin{tabular}{ >{\hspace{-1mm}}l<{\hspace{-1mm}} r r r r r r r r r r }
		\toprule
		& DE & EL & SP & FI & FR & HU & JP & NL & RU & ZH \\
			
		\midrule
		CSS & 15.4 & 3.5 & 20.9 & 9.7 & 16.9 & 9.5 & 14.3 & 11.7 & 17.7 & 5.6 \\
		CV & 4.8  & \it N/A & \it N/A  & \it N/A & 3.0  & \it N/A & \it N/A  & 1.3  & 3.4  & 1.0 \\
		\bottomrule
	\end{tabular}
	
\end{table}

\subsection{CSS10}

CSS10 consists of mono-speaker data in German, Greek, Spanish, Finnish, French, Hungarian, Japanese, Dutch, Russian, and Chinese. It was created from audiobooks and contains various punctuation styles. We applied an automated cleaning to normalize transcripts across languages, including punctuation and some spelling variants (e.g., “œ” $\to$ “oe”).
We romanized Japanese with MeCab and Romkan \cite{mecab, romkan}, Chinese using Pinyin \cite{pinyin}.

\begin{table*}[!th]
	\caption{Left: CERs of ground-truth recordings (\cosi{GT}\xspace) and recordings produced by monolingual and the three examined multilingual models. Right: CERs of the recordings synthesized by \gen and \sha trained on just 600 or 900 training examples per language. Best results for the given language are shown in bold; “*” denotes statistical significance (established using paired t-test; $p<0.05$).}
	\label{tab:cer}
	
	\centering
	\setlength{\tabcolsep}{3pt}
	\setlength{\extrarowheight}{0.5pt}
	\renewcommand{\arraystretch}{0.9}
	\newcommand{\f}{\scalebox{1.0}[1.0]}
	\newcommand{\fb}[1]{\scalebox{1.0}[1.0]{\bm{#1}}}
	\newcommand{\s}[1]{\scalebox{1.0}[1.0]{*\bm{#1}\phantom{*}}}
	\newcommand{\h}[1]{\textbf{#1}}
	\newcommand{\pz}{\phantom{0}} 
	\begin{tabularx}{\textwidth}{l|ccccc<{\hspace{-1mm}}|ccc<{\hspace{-1mm}}>{\hspace{-1mm}}c}
		\toprule
		\textbf{} 		   &
        \h{GT}     & 
		\h{\sgl}    & 
		\h{\sha}    &
		\h{\sep}  & 
		\h{\gen}  & 
		\h{\sha 600}  &
		\h{\sha 900}  &
		\h{\gen 600}  &
		\h{\gen 900}  \\
		\midrule
		DE	& $4.8\pm4.6$ & $7.3\pm6.0$ & $8.3\pm6.0$ & $15.3\pm6.0$\pz & \s{$5.8\pm5.3$} & $13.2\pm8.9\pz$ & \fb{$12.4\pm8.0$}\pz & $15.6\pm9.4\pz$ & $12.5\pm9.3\pz$ \\
		
		EL 	& $8.7\pm6.9$ & \it N/A & \fb{$11.4\pm 8.3$}\pz & $22.2\pm8.3\pz$ & $11.6\pm7.1\pz$ & $16.8\pm9.7\pz$ & $16.0\pm10.2$ & \fb{$14.2\pm8.7$}\pz & $14.7\pm9.8\pz$ \\
		
		SP 	& $3.9\pm4.6$ & $\pz7.0\pm10.8$ & $7.2\pm6.5$ & $10.2\pm8.1\pz$ & \fb{$7.0\pm9.8$} & $9.8\pm7.5$ 	& $9.9\pm8.4$ 	& $ 8.1\pm6.0$ 	& \s{$7.6\pm5.9$} \\
		
		FI 	& $\pz6.9\pm10.4$  & \f{$18.6\pm12.6$} & \fb{$10.3\pm8.0$}\pz & \f{$18.1\pm11.4$} & $10.4\pm7.0\pz$ & $18.2\pm12.2$ & $18.4\pm13.2$ & \s{$13.2\pm10.9$} & $14.0\pm10.6$ \\
		
		FR 	& $11.2\pm7.3\pz$  & \f{$25.2\pm12.6$} & \f{$30.0\pm14.3$} & \f{$54.5\pm21.9$} & \s{$19.0\pm12.9$} & $40.2\pm15.8$ & $37.6\pm16.2$ & $32.9\pm13.2$ & \s{$27.2\pm12.2$} \\
		
		HU	& $6.3\pm6.1$   & $15.8\pm9.5\pz$ & \f{$15.9\pm10.6$} & $18.8\pm9.9\pz$ & \s{$13.5\pm8.3$}\pz & $21.4\pm10.4$ & $21.3\pm13.0$ & \s{$16.5\pm10.4$} & $18.0\pm10.4$ \\  
		
		JP	& $19.0\pm9.3\pz$  & \f{$28.8\pm11.3$} & \f{$27.2\pm11.8$} & \f{$33.7\pm13.5$} & \fb{$25.1\pm12.2$} & $32.5\pm12.8$ & $32.2\pm15.0$ & \fb{$29.9\pm13.0$} & $30.9\pm13.5$ \\ 
		
		NL 	& \f{$14.5\pm7.4\pz$}  & \f{$33.4\pm13.8$} & \f{$31.6\pm12.5$} & \f{$49.0\pm17.4$} & \s{$22.6\pm9.6$}\pz & $37.8\pm13.5$ & $30.4\pm10.2$ & $32.8\pm12.3$ & \fb{$28.3\pm9.8$}\pz \\
		
		RU	& \f{$12.3\pm15.0$} & \f{$45.5\pm24.1$} & \f{$44.4\pm21.9$} & \f{$58.1\pm24.7$} & \s{$34.5\pm21.3$} & $60.4\pm18.6$ & $47.0\pm20.5$ & $38.5\pm20.1$ & \s{$34.4\pm17.9$} \\
		
		ZH	& \f{$14.6\pm11.8$} & \f{$62.8\pm18.5$} & \f{$28.6\pm15.9$} & \f{$27.3\pm14.8$} & \s{$20.5\pm13.6$} & $40.2\pm15.2$ & $39.8\pm18.8$ & $33.0\pm15.5$ & \s{$28.4\pm15.6$} \\
		
		\bottomrule
	\end{tabularx}
\end{table*}

We further filtered the data to remove any potentially problematic transcripts: we preserved just examples
with 0.5-10.1s of audio and 3-190 transcript characters.
We computed means $\mu$ and variances $\sigma$ of audio durations of groups corresponding to examples with the same transcript lengths. Then we removed those with durations outside the interval $(\mu - 3 \sigma, \mu + 3\sigma)$. In total, the resulting dataset includes 125.26 hours of recordings.

\subsection{Common Voice}

To train code-switching models, multi-speaker data is required to disentangle the connection between languages and speakers.
We thus enhanced CSS10 with data from Common Voice (CV) for languages included in both sets -- the intersection covers German, French, Chinese, Dutch, Russian, Japanese, and Spanish. 

Since CV is mainly aimed at speech recognition and rather noisy, we performed extensive filtering:
We removed recordings with a negative rating (as provided by CV for each example) and excluded any speakers with less than 50 recordings. 
We checked a sample of recordings for each speaker, and we removed all their data if we considered the sample to have poor quality. 
This resulted in a small dataset of 39~German, 22~French, 11~Dutch, 6~Chinese, and 6~Russian speakers. Japanese and Spanish data were removed completely. A lot of recordings in CV contain artifacts at the beginning or end. Thus we semi-automatically cleaned leading and trailing segments of all recordings. The dataset has 13.7~hours of audio data in total.

\section{Experiments}

We compare our models described in Section~\ref{sec:models}. The experiment in Section~\ref{sec:exp-training} was designed to show stability and ability to train on lower amounts of data. We conclude that character error rate (CER) evaluation \cite{soukoreff01} is sufficient for this experiment. In Section~\ref{sec:exp-switching}, we test pronunciation accuracy and voice quality of code-switching synthesis. We used a subjective evaluation test as there are no straightforward objective metrics for this task.

We used the same vocoder for all models, i.e., the WaveRNN model trained on a training subset of the cleaned CSS10 dataset.

\subsection{Multilingual training}
\label{sec:exp-training}

\paragraph*{Training setup:}
We used our cleaned CSS10 dataset for training; 64 randomly selected samples per language were reserved for validation and another 64 for testing. We did not have an ambition to clone voices in this experiment, so we switched off speaker classifiers for \sha and \gen (i.e., \sha was reduced to the vanilla Tacotron~2 model with a language embedding).

\begingroup
\setlength{\thickmuskip}{0mu}

We trained the three models for 50k steps with the Adam optimizer.\footnote{With $\beta_1 = 0.9$, $\beta_2 = 0.999$, $\epsilon = 10^{-6}$, and weight decay of $10^{-6}$} We used a stepped learning rate that starts from $10^{-3}$ and halves every 10k steps. In the case of \sep, we used a lower initial learning rate $10^{-4}$. For \sgl, the learning rate schedule was tuned individually per language. We stopped training early after validation data loss started increasing. \sha, \sep, and \gen used speaker embeddings of size 32 and \gen 
used language embeddings and parameter generators of size 10 and 8, respectively. We used language-balanced batches of size 60 for all models.

\endgroup

\paragraph*{Evaluation:}

We synthesized evaluation data using all the models followed by WaveRNN and we sent the synthesized recordings to Google Cloud Platform ASR.\footnote{\url{https://cloud.google.com/speech-to-text}} Then we computed CERs between ground-truth and ASR-produced transcripts (we used the native symbols for Chinese and Japanese).


\paragraph*{Results:}
Table~\ref{tab:cer} summarizes the obtained CERs. The first column gives us a notion about the performance of the ASR engine. The rates stay below 20\% for all languages; higher CERs are mostly caused by noisy CSS10 recordings.

We were not able to train the Greek \sgl model due to low amount of training data. The decoder started to overfit soon before the attention could have been established. The performance of \sgl is similar to \sha except for Chinese, Finnish, and Greek. \sep performed noticeably worse than \sha or even \sgl. This may be caused by the imbalance between the batch size of the encoder and the decoder as the encoder's effective batch size is just $B/L$.\footnote{Our attempts to compensate for this using different encoder and decoder learning rates were not successful.} Sharing of the data probably regularized the decoder, so the attention was established even in the case of Greek. \gen seems to be significantly better than \sha on most languages. It fulfills our expectations as \gen should be more flexible. 

\paragraph*{Manual error analysis:} We manually inspected the outputs in German, French, Spanish, and Russian. In the case of Spanish, all the models work well; we noticed just differences in the treatment of punctuation. German outputs by \gen seem to be the best. Other models sometimes do unnatural pauses when reaching a punctuation mark. Right after the pauses, they often skip a few words. \gen is noticeably better on French and Russian, others produce obvious mispronunciations.

\paragraph*{Data-stress training:} To further test the models in data-stress situations, we chose random subsets of 600 and 900 examples per language from the training set (i.e., about 80 or 120 minutes of recordings, respectively). 
We trained all models on both reduced datasets, but 
accomplished training just for \sha and \gen. 
While training on the bigger and smaller dataset, we decayed the learning rate every 7.5k and 5k training steps, respectively.
The right half of Table~\ref{tab:cer} shows that \gen can work better even in data-stress situations. \gen models have, compared to \sha models, significantly better CER values on six languages.

\subsection{Code-switching}
\label{sec:exp-switching}

\begingroup
\setlength{\thickmuskip}{0mu}

\paragraph*{Training setup:}
In this experiment, we only used the five languages where both CSS10 and CV data are available (Table~\ref{tab:data}), and trained on all data in our cleaned sets; 64 and 4 randomly selected samples for each speaker from CSS10 and CV, respectively, were reserved for validation. The \sgl models are not applicable to the code-switching scenario. \sha, \sep, and \gen models were trained for 50k steps with the same learning rate and schedule settings as in Section~\ref{sec:exp-training}, this time with the adversarial speaker classifiers enabled.\footnote{Based on preliminary experiments on validation data, we set $\lambda = 1$ and weighted the loss of the classifier by $0.125$ and $0.5$ for \gen and \sha, respectively. The classifiers include a hidden layer of size 256.} We set the size of speaker embeddings to 32 and used a language embedding of size 4 in \sha. \gen uses language embeddings of size 10 and generator layers of size 4. We used mini-batches of size 50 for all models.

\endgroup

\begin{figure}[t]
	\centering
	\includegraphics[width=\linewidth]{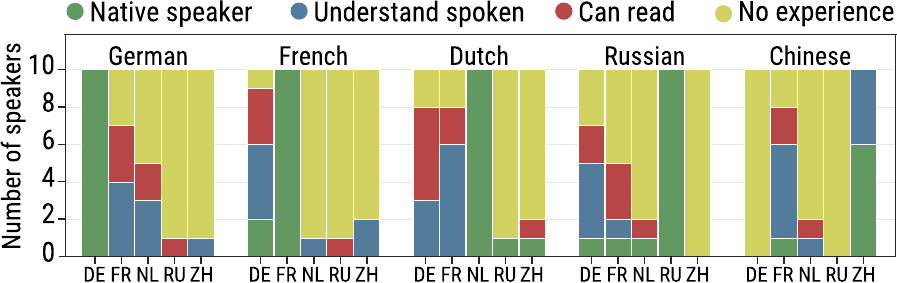}
	\caption{Language abilities of participants of our survey. 
	}
	\label{fig:demo}
\end{figure}

\begin{figure}[ht]
	\centering
	\includegraphics[width=\linewidth]{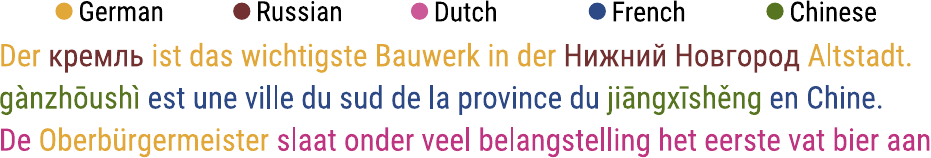}
	\caption{Examples of code-switching evaluation sentences.}
	\label{fig:codeswitch}
\end{figure}

\paragraph*{Code-switching evaluation dataset:}
We created a new small-scale dataset especially for code-switching evaluation. We used bilingual sentences scraped from Wikipedia. For each language, we picked 80 sentences with a few foreign words (20 sentences for each of the 4 other languages); Chinese was romanized. We replaced foreign names with their native forms (see Fig.~\ref{fig:codeswitch}). 

\paragraph*{Subjective evaluation:}
We synthesized all evaluation sentences using speaker embedding of the CSS10 speaker for the base language of the sentence.
We arranged a subjective evaluation test and used a rating method that combines five-point mean opinion score (MOS) with MUSHRA \cite{itu_mushra}.
For each sample, its transcript and systems' outputs were shown at the same time. Participants were asked to rate them on a scale from 1 to 5 with 0.1 increments and with labels “Bad”, “Poor”, “Fair”, “Good”, “Excellent”. 
To distinguish different error types, we asked for two ratings: (1) \emph{fluency}, naturalness, and stability of the voice (speaker similarity) -- to check if foreign words cause any change to the speaker's voice, and (2) \emph{accuracy} -- testing if all words are pronounced and the foreign word pronunciation is correct.
Participants could leave a textual note at the end of the survey.

For each language, we recruited ten native speakers that spoke at least one other language fluently via the Prolific platform (Fig.~\ref{fig:demo}).\footnote{\url{https://www.prolific.co}; 4 participants who reported as Chinese native speakers on Prolific only reported non-native fluency in our survey.} 
They were given twelve sentences with the base language matching their native language where each of the other languages was represented by three sentences.\footnote{In 3 sentences, a random model output was distorted and used as sanity check (expected to be rated lowest). All participants passed.}

\begin{figure}[h]
	\centering
	\includegraphics[width=\linewidth]{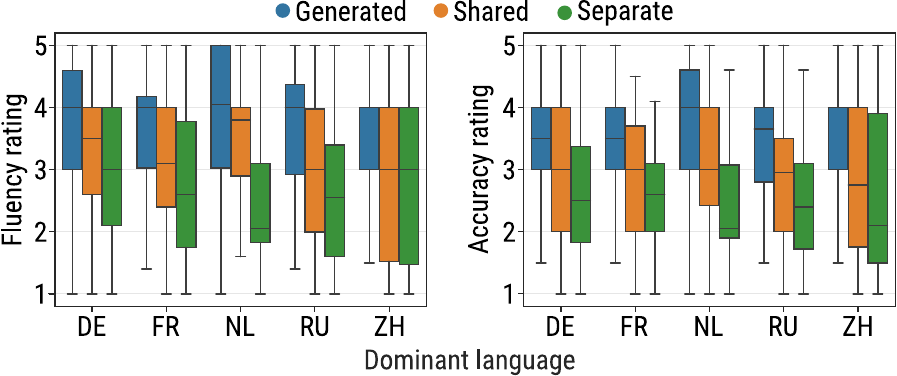}
	\caption{Graphs showing distributions of fluency and accuracy ratings grouped by the dominant language of rated sentences.}
	\label{fig:aaa}
\end{figure}

\paragraph*{Results:}
Table~\ref{tab:mos} summarizes results of the survey. The rows marked “All” show means and variances of the ratings of all 50 participants. Fig.~\ref{fig:aaa} visualizes quantiles of the ratings (grouped by dominant languages). \gen has significantly higher mean ratings on both scales. Unlike \sha or \sep, it allows cross-lingual mixing of the encoder outputs and enables smooth control over pronunciation.
\sep scores consistently worst. The accuracy ratings are overall slightly higher than the fluency ratings; this might be caused by improper word stress, which several participants commented on.

\begin{table}[t]
	\caption{Mean (with std.~dev.) ratings of fluency, naturalness, voice stability (top) and pronunciation accuracy (middle). The bottom row shows the number of sentences with word skips.
	\label{tab:mos}
	}
	\centering
	
	\setlength{\extrarowheight}{0pt}
	\renewcommand{\arraystretch}{0.9}
	\renewcommand{\aboverulesep}{0.05pt}
	\renewcommand{\belowrulesep}{2.5pt}
	
	\begin{tabular}{ l c c c c }
		\toprule
		& &
		\textbf{\sha}  & 
		\textbf{\sep} &
		\textbf{\gen} \\	
		\midrule
		
		\multirow{6}{*}{\rotatebox[origin=c]{90}{\Tstrut\bf Fluency}}
		& German 
			& $3.0\pm1.1$ & $2.6\pm1.0$ & *\bm{$3.4\pm0.9$} \\ 
		& French 
			& $2.8\pm1.0$ & $2.6\pm1.0$ & *\bm{$3.5\pm0.9$} \\ 
		& Dutch 
			& $3.1\pm0.9$ & $2.5\pm1.1$ & *\bm{$3.7\pm1.0$} \\ 
		& Russian 
			& $2.8\pm1.0$ & $2.5\pm1.0$ & *\bm{$3.4\pm0.9$} \\ 
		& Chinese
			& $2.7\pm1.3$ & $2.6\pm1.2$ & *\bm{$3.5\pm1.2$} \\
		& \textbf{All}
			& $2.9\pm1.1$ & $2.5\pm1.1$ & *\bm{$3.5\pm1.0$} \\
		
		\midrule
		
		\multirow{6}{*}{\rotatebox[origin=c]{90}{\Tstrut\bf Accuracy}}
		& German 
		 & $3.3\pm1.1$ & $3.1\pm1.2$ & *\bm{$3.7\pm1.0$} \\
		& French 
		 & $3.1\pm1.1$ & $2.7\pm1.2$ & *\bm{$3.7\pm0.9$} \\
		& Dutch
		 & $3.4\pm1.0$ & $2.5\pm1.2$ & *\bm{$3.9\pm1.1$} \\
		& Russian 
		 & $3.0\pm1.2$ & $2.6\pm1.2$ & *\bm{$3.6\pm1.0$} \\
		& Chinese
		 & $2.9\pm1.4$ & $2.8\pm1.4$ & *\bm{$3.5\pm1.2$} \\
		& \textbf{All}
		 & $3.1\pm1.2$ & $2.7\pm1.2$ & *\bm{$3.7\pm1.1$}  \\
		
		\midrule
		\multicolumn{2}{c}{\textbf{Word skips}} & 41/400 & 38/400 & \textbf{11/400} \\
		\bottomrule
	\end{tabular}
	
\end{table}

\paragraph*{Manual error analysis:} 
We found that the models sometimes skip words, especially when reaching foreign words in Chinese sentences.
Therefore, we manually inspected all 400 outputs of all models and counted sentences where any word skip occurred,
see the “Word skips” row in Table~3. We found that the \gen model makes much fewer of these errors than \sha and \sep.

\section{Conclusion}

We presented a new grapheme-based model that uses meta-learning for multilingual TTS. We showed that it significantly outperforms multiple strong baselines on two tasks: data-stress training and code-switching, where our model was favored in both voice fluency as well as pronunciation accuracy.
Our code is available on GitHub.\textsuperscript{\ref{fn:github}}
For future work, we consider changes to our model's attention module to further improve accuracy.

\section{Acknowledgements}

This research was supported by the Charles University grant PRIMUS/19/SCI/10. 

\clearpage

\renewcommand\bibsection{\section{References}}
\bibliographystyle{IEEEtranN}
\bibliography{template}

\end{document}